# Trade-off between angular resolution and straylight contamination in CMB anisotropy experiments

## I. Pattern simulations

M. Sandri[1], F. Villa[1], R. Nesti[2], C. Burigana[1], M. Bersanelli[3], and N. Mandolesi[1]

[1] IASF – CNR, Sezione di Bologna, Via P. Gobetti, 101, I-40129 Bologna, Italy  
   e-mail: `sandri, villa, burigana, mandolesi @bo.iasf.cnr.it`
[2] INAF – Osservatorio Astrofisico di Arcetri, L.go Fermi, 5, Firenze  
   e-mail: `nesti@arcetri.astro.it`
[3] Dipartimento di Fisica, Universitá degli Studi di Milano, Via Celoria, 16, I-20133 Milano, Italy  
   e-mail: `bersanelli@uni.mi.astro.it`

**Abstract.**
The study of Cosmic Microwave Background (CMB) anisotropies represents one of the most powerful Cosmological tools. After the great success of the two NASA's satellite missions COBE and WMAP, PLANCK represents the third generation of mm-wave instruments designed for space observations of CMB anisotropies within the new Cosmic Vision 2020 ESA Science Programme. The PLANCK survey will cover the whole sky with unprecedented sensitivity, angular resolution and frequency coverage, using two instruments that share the focal region of a 1.5 m off-axis dual reflector telescope: the Low Frequency Instrument (LFI) and the High Frequency Instrument (HFI). Within the LFI optical interfaces optimisation activity, two concurrent demands have to be satisfied: the best angular resolution (which impacts the ability to reconstruct the anisotropy power spectrum of the CMB anisotropies at high multipoles) and the lowest level of straylight contamination (that may be one of the most critical sources of systematic effects). We present the results of the optical simulations aimed to establish the trade-off between angular resolution and straylight rejection, carried out for the 100 GHz channel of PLANCK Low Frequency Instrument. Antenna pattern of different models of dual profiled corrugated conical feed horns have been simulated using advanced simulation techniques, considering the whole spacecraft geometry in order to obtain truthful sidelobe predictions. Optical computation accuracy necessary to provide strong straylight evaluation in reasonable computational time is shown and the inadequacy of a Gaussian feed model in realistic far pattern predictions is demonstrated. This paper is based on LFI activities.

## 1. Introduction

The study of Cosmic Microwave Background (CMB) anisotropies represents one of the most powerful Cosmological tools. The shape of the angular power spectrum of CMB anisotropies sensitively depends on fundamental cosmological parameters, so that an accurate measure of the spectrum provides a unique method to establish the value of these parameters with high precision. After the discovery of the CMB anisotropies by the NASA satellite COBE (Smoot et al. 1992), several ground-based and balloon-borne experiments have been set up with the purpose to measure the anisotropies at sub-degree angular scales, by using dedicated reflecting telescopes and interferometer techniques (see e.g. Bersanelli et al. 2003 for a review). Recently the NASA satellite mission WMAP (Wilkinson Microwave Anisotropy Probe) equipped with a pair of 1.4 m back-to-back telescopes (Page et al. 2003a) reconstructed the power spectrum with high precision at multipoles up to ∼ 800 (Bennett et al. 2003, Page et al. 2003b). Within the Horizon 2000 Scientific Programme, ESA planned PLANCK as the third generation of mm-wave instruments designed for space observations of CMB anisotropies. PLANCK will be launched in 2007 and will carry the state-of-the-art of microwave radiometers (Low Frequency Instrument, Mandolesi et al. 1998) and bolometers (High Frequency Instrument, Puget et al. 1998) coupled with a 1.5 m telescope and working between 30 and 900 GHz in nine frequency channels.

Owing to the small amplitude of the CMB anisotropies (about $10 \div 100 \, \mu$K rms), the accurate control of systematic effects is mandatory to perform high precision CMB measurements, with experiments either from space or on ground. Specifically the optics (composed by the telescope–feed horns assembly) is one of the major limiting factor for high precision CMB measurements, being aberrations of the main beam and straylight two main sources of systematic errors. The angular resolution, eventually degraded by the optical aberrations, limits the reconstruction of the anisotropy power spectrum at high multipoles (Mandolesi et al. 2000). Otherwise, the fluctuations of the straylight signal (Burigana et al. 2001) contaminate the measurements mainly at large and intermediate angular scales (i.e. at multipoles $\ell$ less than $\approx 100$), and must be kept below a level of few $\mu$K.

Accurate predictions and measurements of the antenna pattern are an essential requisite both during the instrument development phase and for an in-depth knowledge of the whole instrument response in the development of the data reduction pipeline, for all high precision CMB experiments.

Robust optical simulations are of primary importance in the understanding of the straylight rejection capability of the instrument and telescope, in particular for the far sidelobes region (at large angles from the main beam), where the power levels are extremely low and direct measurements become difficult and uncertain,

The antenna response features at large angles from the beam centre are determined mainly by diffraction and scattering from the edges of the mirrors and from nearby supporting structures. Therefore they can be reduced by decreasing the illumination at the edge of the primary mirror, i.e. by increasing the Edge Taper (ET), defined as the ratio of the power per unit area incident on the centre of the mirror to that incident on the edge[1]. Of course, the higher is the ET, the lower is the sidelobe level and the straylight contamination. On the other hand, increasing the ET has a negative impact on the angular resolution, which reduces the ability to reconstruct the anisotropy power spectrum of the CMB anisotropies at high multipoles. With a given telescope the ET can be varied by changing the feed horn design, since the feed horn pattern determines the illumination function of the telescope. Clearly, an appropriate choice of the feed design will impact on the angular resolution and on the straylight rejection as well.

In the framework of the PLANCK Low Frequency Instrument (LFI) optical interface optimisation, the trade-off between angular resolution and straylight rejection has been obtained. The work is presented in this paper, Paper I, and in a joined paper (Burigana et al. 2003, hereafter Paper II). Three steps have been followed.

1. Computation of realistic feed horn pattern. Electromagnetic design of feed horns have been performed in order to simulate realistic radiation patterns with different edge tapers.
2. Computation of the full optical response. Each feed horn pattern has been propagated through the PLANCK optical system to obtain the complete $4\pi$ beam. Although several powerful electromagnetic simulation methods, like Phsical Optics (PO), Physical Theory of Diffraction (PTD), Geometrical Optics (GO), and Geometrical Theory of Diffraction (GTD), are well understood and widely used, their applicability to real optical systems can be very difficult particularly when many reflecting structures have to be simulated. Multiple diffractions and reflections between optical elements (reflectors, one or more shields, supporting structures) have to be considered and this leads to unacceptable computational time. To overcome this difficulty, an advanced simulation technique has been used and it is described in detail in this paper.
3. Straylight evaluation. This has been calculated by convolving the full pattern with the sky signal by considering the observational strategy. This step is described in Paper II.

We present, as a working case, the optimisation activity carried out on two representative feed horns of the PLANCK LFI, aimed to reach the best angular resolution in line with straylight requirements. By means of accurate radiation pattern simulations, the details of the antenna response have been computed for four different models of the feed horn #4 and three models of the feed horn #9 (see the right panel of Fig. 1). Through the comparison between the level of Galactic straylight contamination, described in Paper II, the best horn design has been identified.

We are currently applying the optimisation procedure presented here to the other LFI frequency channels but the method and basic results reported in this work can be in principle applied also to different experiments aimed to measure CMB anisotropies. Specifically, the assessment of the optical computation accuracy necessary to provide robust straylight evaluation – in reasonable computational time – is a basis of optimisation procedures of each CMB anisotropy experiment.

In Sect. 2 the PLANCK optical configuration (telescope, shields and focal plane unit) is described. In Sect. 3 the electromagnetic design (corrugation profiles and far field pattern simulations) of the feed horns analysed are presented, as well as major electromagnetic characteristics of each model. Sect. 4 reports a description of simulation methods used to compute far field patterns. In Sect. 5 the simulation results are reported, divided in main beam pattern, intermediate pattern, and far field pattern. The correlation between the computed integrated power, normalized to the total power, from far sidelobes (and intermediate beam) and the linear ET (LET, i.e. $10^{-ET/10}$) is reported in Sect. 6, as well as the correlation between the Galactic straylight induced noise (antenna temperature in $\mu$K, rms and peak-to-peak values) and the LET, both relevant in future optimisation activities. Finally, conclusions are reported in Sect. 7. In Appendix A some useful quantity related to main beams are reported, whereas Appendix B reports the computed beam patterns not included in Sect. 5.3.

---

[1] The ET is not constant but varies along the reflector rim. For convenience we define the ET at feed level as a dB level below the peak at certain angle from the feed boresight (dB @ angle)

## 2. Electromagnetic model of the optics

Accurate pattern simulations require a realistic representation of the optics in the electromagnetic model. Although for the main beam simulations only the telescope geometry can be considered, for the intermediate and far angles pattern also the structures surrounding the reflecting mirrors needs to be considered in detail. These structures could be the mirror mounting structure, the shields and, in a general context, all the surfaces could emit, reflect, or scatter radiation from the sky to the feed (i.e. detectors) (Mennella et al. 2002).

In the case of the PLANCK spacecraft, the geometry can be divided in four groups (see Fig. 1): the dual reflector telescope, the telescope main baffle (a shield surrounding the reflectors), the cryo-structure (three V– grooves under the telescope), and the Service Module, at the bottom of the satellite. However, the PLANCK optical environment is dominated by a large baffle surrounding the telescope and, on the bottom, the higher of the three thermal shields (V– grooves) used to thermally decouple the telescope and the focal assembly from the service module (Villa et al. 2002b). Then, for the simulations reported in this paper only the telescope, the baffle and the first V– groove have been considered. The Service Module and the other two V– grooves could contribute to the pattern after the diffraction on the first V– groove which then dominates the pattern at large angles and low levels ($< -100$dB).

The PLANCK Telescope (Dubruel et al. 2000, Villa et al. 2002a) is designed as an off-axis tilted system offering the advantage to accomodate instruments in a large focal surface with an unblocked aperture, maintaining the diffraction by the secondary mirror and supporting structures at very low levels. The baseline configuration has been selected among several different optical designs, during the PLANCK Payload Architect industrial activity, in the 1999. The current configuration has been obtained by optimising the telescope performance for a set of equally distributed in frequency (from 30 to 857 GHz) and space (within the focal plane box) representative feed horns. Both mirrors have an ellipsoidal shape (aplanatic configuration). The conical constants, the focal length, the tilting, and the decentre of the mirrors have been combined to reduce the main beam aberrations, the curvature of the focal surface, and the spillover as well. The primary mirror physical dimensions are about $1.9 \times 1.5$ meters, allowing a projected circular aperture of 1.5 meter of diameter. The secondary reflector has been oversized up to approximately 1 meter of diameter to avoid an under illumination of the primary. The sub reflector revolution axis is tilted with respect to the main reflector revolution axis of about $10°$. The telescope field of view is $\pm 5°$ centred on the line of sight (LOS) which is tilted at about $3.7°$ with respect to the main reflector axis, and forms an angle of $85°$ with the satellite spin axis, typically oriented in the anti-Sun direction during the survey.

The telescope is surrounded by the so-called main baffle that protects it from straylight and provides the optimum radiative surface that passively cools the telescope at approximately 50 K. The baffle is interfaced with the coldest of three V– groove shields, located at the bottom of the telescope, that ensure the passive thermal control design (see Fig. 1).

## 3. The focal plane unit and the feed horns

The Low Frequency Instrument is coupled to the PLANCK telescope by an array of corrugated feed horns. Dual profiled corrugated horns have been selected as the best design in terms of shape of the main lobe, very low level of cross polarisation, level of sidelobes, control of the phase centre location, low weight and compactness (Clarricoats & Olver 1984; Olver & Xiang 1988; Villa et al. 2002c).

The focal plane layout considered in this work foresees 12 feed horns at 100 GHz, 6 at 70 GHz, 3 at 44 GHz and 2 at 30 GHz, distributed around the HFI, as reported in Fig. 1. In this study our attention is focussed on the feed horns #4 and #9 (hereafter, LFI4 and LFI9), both at 100 GHz. Analogous considerations hold for the feed horns #10 and #15, located symmetrically in the opposite side of the Focal Plane Unit (FPU).

Four models of the LFI4 (4A, 4B, 4C, 4D) and three models of the LFI9 (9A, 9B, and 9C) have been analyzed. In Fig. 2 and Fig. 3 the corrugation geometries and the radiation patterns are shown. The seven different patterns have been computed by Modal Matching/MoM models on dual profile corrugated horns in 72 azimuthally equidistant cuts, in which $\theta$ (angle from the boresight) ranges from $0°$ to $180°$.

4A (ET 28.3 dB @ $24°$) and 9A (ET 25.5 dB @ $24°$) represent two models, for the LFI4 and LFI9, whose edge taper values ensure a good straylight rejection (lower than 2 $\mu$K peak-to-peak), to the detriment of the angular resolution (12.85 ′ and 10.53′ for the models 4A and 9A, respectively). Other three different designs have been performed for the LFI4 (4B, 4C, and 4D) with the same edge taper (ET 19.0 dB @ $24°$) that should lead to an angular resolution of about 12′. Since it has been ascertained that a further edge taper degradation doesn't lead to an angular resolution improvement for this horn, because of the strong illumination of the mirrors that increase the aberrations on the main beam, we tried to improve the angular resolution changing the horn design, edge taper being equal. In Fig. 2 it should be noted the corrugation profile of these three models and, in particular, the unusual profile of the latter two models, 4C and 4D. Two different designs have been performed for the LFI9: 9B (ET 19.0 dB @ $24°$) and 9C (ET 15.0 dB @ $24°$), that lead to angular resolutions of 10′ and 9.5′, respectively. In Fig. 3 it should be noted that an edge taper degradation involve a smaller feed horn aperture. Astrophysical simulations, reported in Paper II, based on the beams computed using these feed models coupled with the PLANCK telescope, told us which is the best choice in terms both of angular resolution and straylight contamination, for the two feed horns under study.

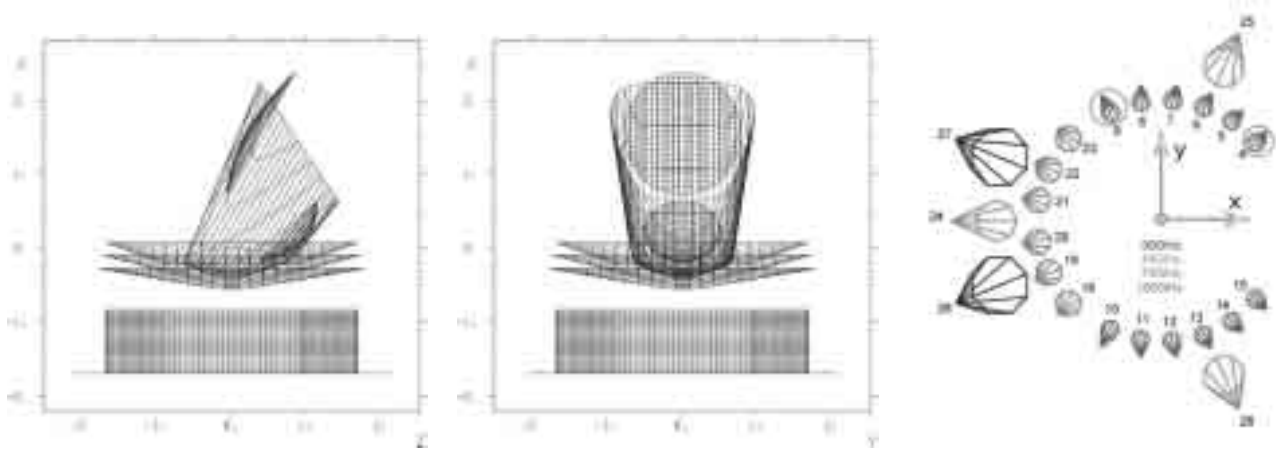

**Fig. 1.** Telescope and shields geometry in $(XZ)_{TEL}$ (*left*) and $(XY)_{TEL}$ (*centre*) planes; Planck/LFI Focal Plane Unit configuration represented in the Reference Detector Plane axis system. The two feed horn under study (LFI4 and LFI9 at 100 GHz) are identified with a circle (*right*).

## 4. Simulation techniques

The simulations are performed by considering the feed as a source and by computing the pattern scattered by both reflectors on the far field using GRASP8[2]. To predict the radiation pattern, different techniques can be applied: Physical Optics (PO), Physical Theory of Diffraction (PTD), Geometrical Optics (GO), and Geometrical Theory of Diffraction (GTD). PO is the most accurate method and may be used in all regions of the space surrounding the reflector antenna system. The field of the source is propagated on the reflector to calculate the current distribution on the surface. Then, the currents are used for evaluating the radiated field from the reflector. The calculation of the currents close to the edge of the scatterer are modelled by PTD. Unfortunately, as the frequency increases the reflectors have to be more and more precisely sampled. As a consequence, the density of the integration grids, in which currents have to be computed on the reflectors, must be finer and the computation time becomes huge. For a two-reflectors antenna system like Planck, the computation time increases with the fourth power of the frequency [3]. Moreover, proper evaluation of the effect of shields is crucial for the optimisation of the edge taper, since these structures redistribute the power that is radiated by the horns and is not reflected by the telescope.

Although to predict accurately the antenna pattern of our model of the telescope a full PO computation would be required, this is not the case for the whole spacecraft simulations since the PO approach cannot be correctly applied when many multiple diffractions and reflections between scatterers are involved. For this reason, in our simulations we use a new GRASP8 technique, named Multi-reflector GTD (MrGTD), that computes the scattered field from the reflectors performing a backward ray tracing, and represents a suitable method for predicting the full-sky radiation pattern of complex mm-wavelength optical systems in which the computational time is frequency independent. When many scatterers are involved, the amount of ray tracing contributions may lead to unacceptable computation time even with MrGTD. Therefore, is crucial to identify *all and only* the contributions (i.e. sequences of diffractions and/or reflections on each scatterer, see Sandri et al. (2002) for a detailed description of the method applied to the Planck case) which produce significant power levels in the resulting radiation pattern. We set a threshold equal to $-100$ dB, $-50$ dBi at 100 GHz, since this is the required straylight rejection level in this frequency channel. Lower power levels not produce significant straylight contamination from the diffuse Galactic emission, as reported in Burigana et al. (2001) and how can be inferred from the results presented in this work and in Paper II.

In our simulations we have considered the two reflectors, the baffle and the first V– groove as blocking structures. The simplest ($1^{st}$ order) optical contributions producing significant power levels are reflections on the sub reflector, on the main reflector, and on the baffle, as well as diffractions on the sub reflector, on the main reflector, and on the baffle. Other non-negligible contributions are found considering two interactions with the reflectors ($2^{nd}$ order – for example, rays reflected on the baffle and then diffracted by the main reflector), three interactions ($3^{rd}$ order – for example, rays reflected on the sub reflector, diffracted by the main reflector, and then diffracted by the baffle) and so on. In this framework, we have not considered reflections or diffractions on the V– groove, as well as higher order contributions, since we expect them at very low levels (less than $-100$ dB). In Fig. 4 is shown the ray tracing of some relevant MrGTD contributions: in the left panel are drawn rays coming from the feed not intercepted by the reflecting structures; in the central panel is shown the ray tracing of the $1^{st}$ order optical contribution coming from the rays reflected on the sub reflector falling in the main spillover region with significant power levels (at about $-80$ dB); in the right panel are represented rays reflected on the sub reflector and then diffracted by the main reflector, reaching the main spillover region as well. In Fig. 5 is reported the far-field pattern of LFI9 B in the plane $\phi_{bf}$ = 54°, passing through the main spillover, computed at the $1^{st}$ order, at the $2^{nd}$ order, and up to the $3^{rd}$ order. The optical contribution of the direct rays from the feed is also plotted, together with the pattern of the feed computed without blocking structure. It is evident how in the region with $\theta_{bf}$ between few degrees and 60° feed sidelobes prevail. Therefore, using a Gaussian feed model (accurate enough for the main beam computation) is completely misleading for a realistic far pattern prediction.

---

[2] GRASP8 is a software developed by TICRA (Copenhagen, DK) for analysing general reflector antennas.
[3] Without considering the effect of the shields, this means more than 800 hours to perform a full pattern computation at 100 GHz, using a 550 MHz Pentium II machine.

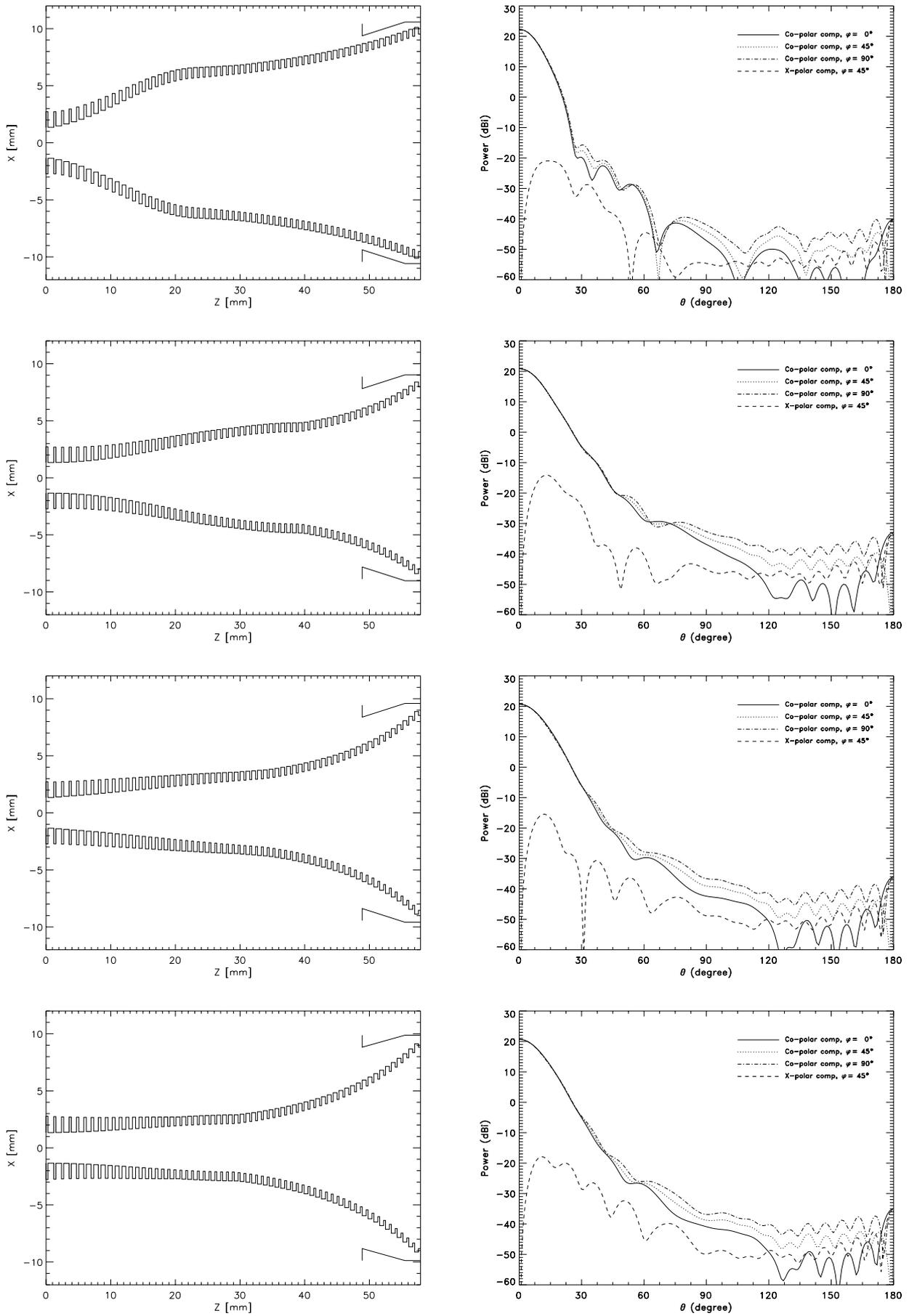

**Fig. 2.** Corrugation profiles of the four models 4A, 4B, 4C, 4D of LFI4 (*left*) and relative beam patterns (co– and x– polarization) in the principal planes E and H, and in the plane with $\phi$ = 45 degrees (*right*).

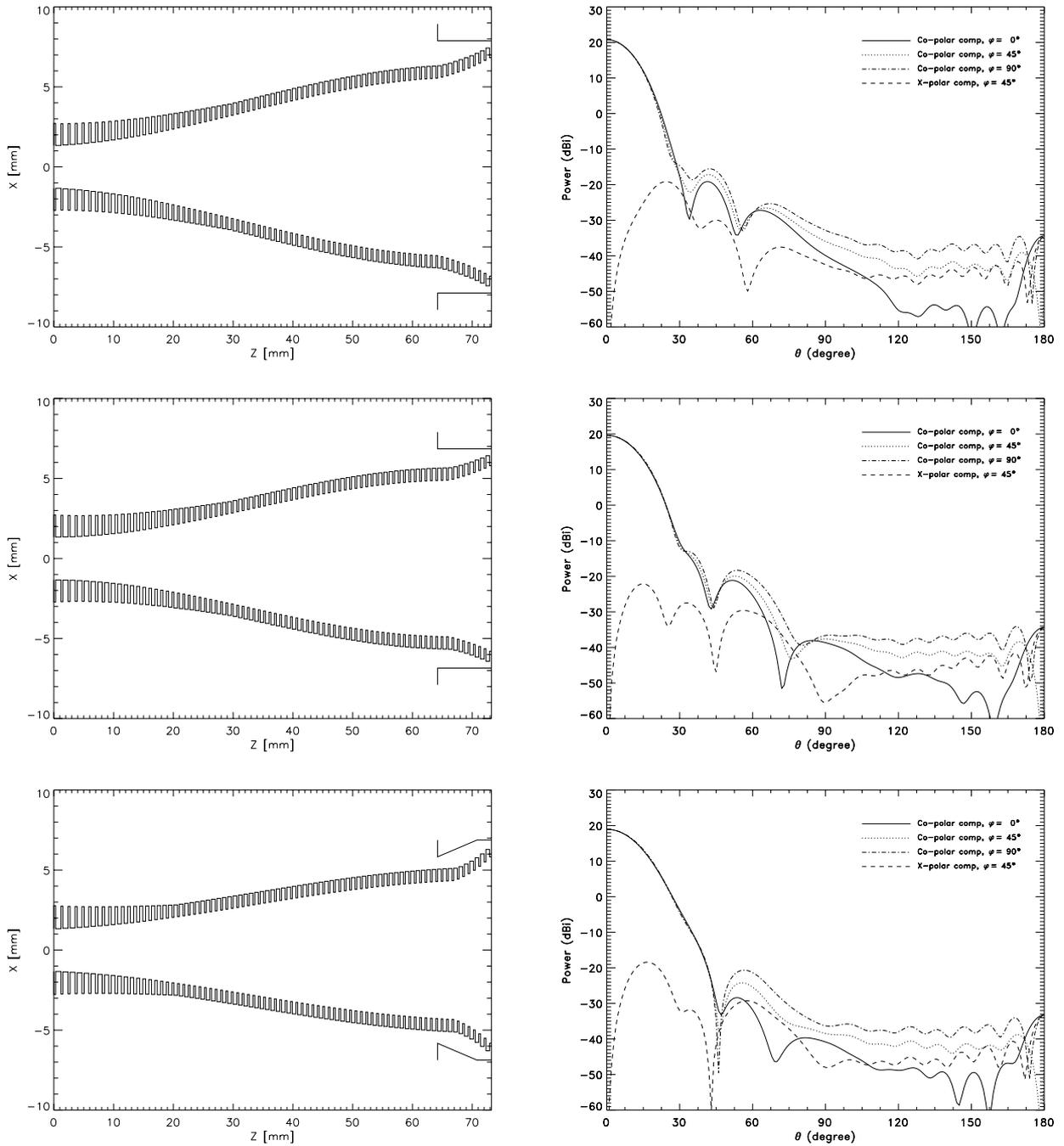

**Fig. 3.** Corrugation profiles of the three models 9A, 9B, 9C of LFI9 (*left*) and relative beam patterns (co– and x– polarization) in the principal planes E and H, and in the plane with $\phi$ = 45 degrees (*right*).

## 5. Full pattern simulations

For each LFI feed horn we have defined a corresponding *beam frame* on the sky, starting from the telescope Line of Sight (LOS) through three angles: $\theta_B$, $\phi_B$, and $\psi_B$. The first two angles are related on the main beam pointing direction, whereas $\psi_B$ defines the main beam polarization direction in a plane perpendicular to the main beam direction. In the beam frame, ($x_{bf}, y_{bf}, z_{bf}$) in Cartesian coordinates or ($\theta_{bf}, \phi_{bf}$) in polar coordinates, the peak of the power pattern falls in the centre of the $(UV)_{bf}$– plane (i.e. along the $z_{bf}$ direction, with $\theta_{bf} = 0$).

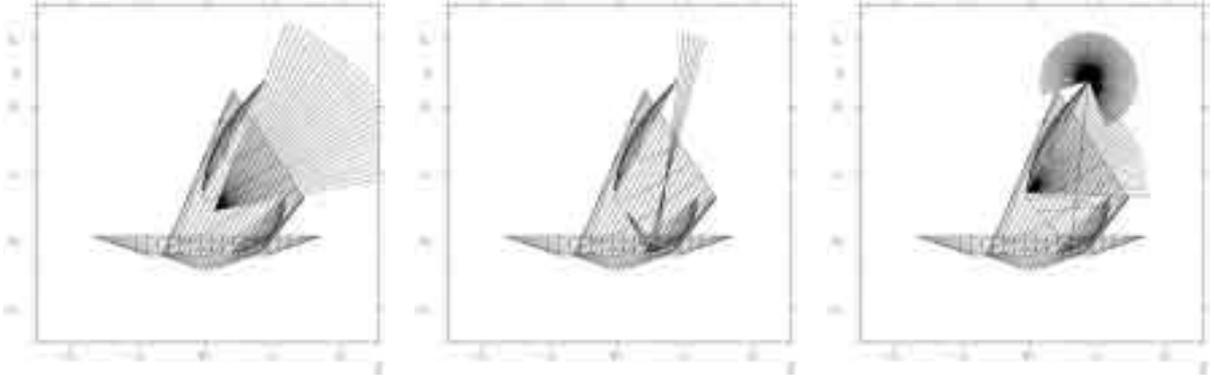

**Fig. 4.** Ray-tracing of some MrGTD contributions in the symmetry plane (for clearly, the source is placed in the centre of the FPU): direct rays from the feed (*left*), rays reflected on the lower part of the sub reflector (*centre*), rays reflected on the sub reflector and then diffracted by the main reflector (*right*).

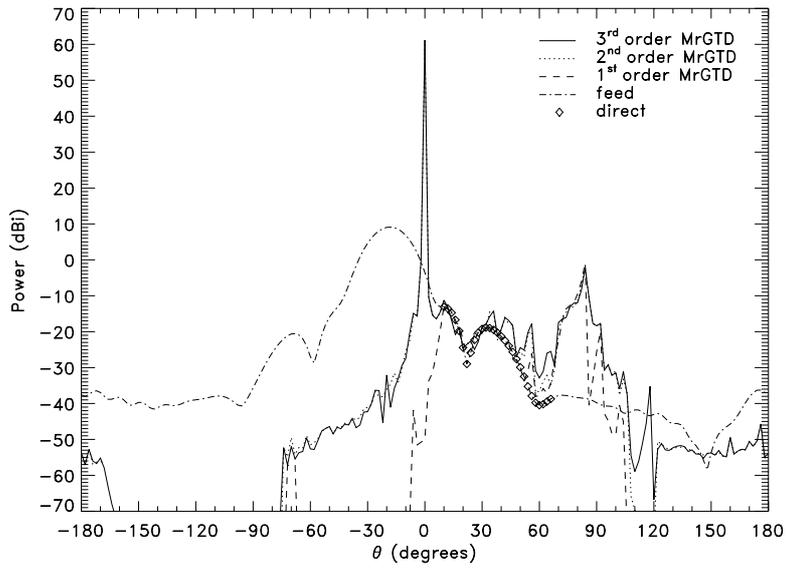

**Fig. 5.** Far-field power pattern of LFI9 B in the plane $\phi_{bf} = 54°$, passing through the main spillover. Differences between simulations at the $1^{st}$, $2^{nd}$, and up to the $3^{rd}$ order are evident. It should be noted the $3^{rd}$ order peak at $\theta_{bf}$ between $110°$ and $120°$. The contribution of the direct rays from the feed is pointed out, and the feed sidelobes appear at about $-20$ dBi, between the main beam direction and the main spillover.

We have divided the far-field pattern in three regions at different angular distance from the beam centre: the main beam region ($\theta_{bf} < 1.2°$), the intermediate beam region ($1.2° < \theta_{bf} < 5°$), and the far beam region ($\theta_{bf} > 5°$). The choice of these angles has been driven by the fact that at LFI frequencies, for the PLANCK telescope, the first minimum of the beam pattern falls at about $1.2°$, whereas $5°$ roughly divides pattern regions where significant response variations occur on angular scales less than $1°$ from those where they occur on degree or larger scales (Burigana et al. 2001). In each of the three regions, simulations are performed using different methods, as explained as follows.

### 5.1. Main beam

The main beam simulations have been performed using PO analysis on each reflector. Far field radiation patterns have been computed in the co– and x– polar basis according to the Ludwig's third definition (Ludwig 1973) in UV– spherical grids with 301x301 points, in order to compute the main beam angular resolution of each feed model analysed, as well as all major characteristics reported in Tab. 1. Contour plots of the main beams computed with the four models of the LFI4 and the three models of the LFI9 are shown in Fig. 6.

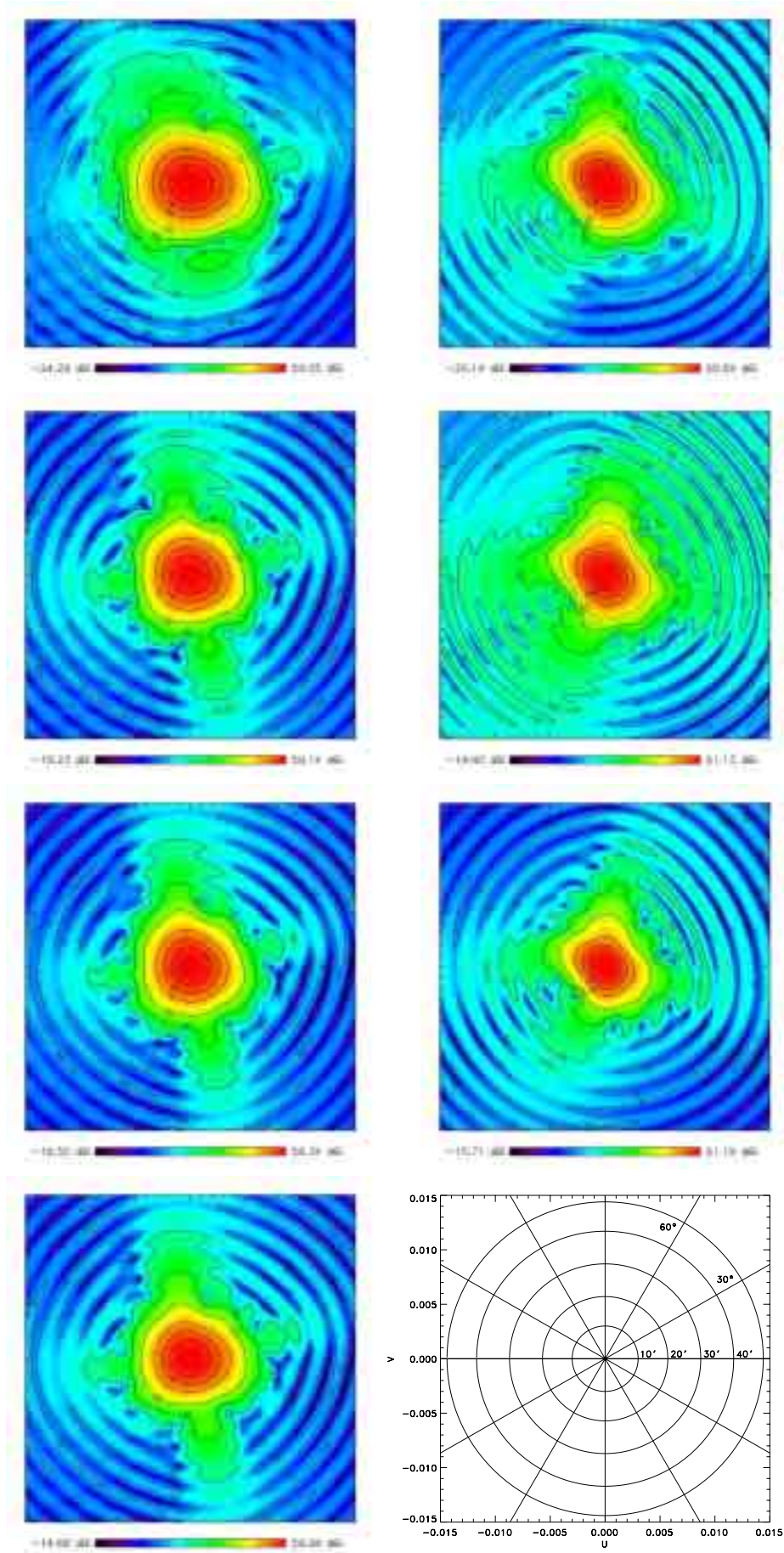

**Fig. 6.** Contour plots in the UV– plane (−0.015 < U,V < 0.015) of the main beam computed for the four models of LFI4 (*left column*, from the top to the bottom: 4A, 4B, 4C, 4D) and for the three models of LFI9 (*right column*, from the top to the bottom: 9A, 9B, 9C). The lines of the contour plots, normalized at peak gain (reported in Tab. 1), are at −3, −6, −10, −20, −30, −40, −50, −60, and −70 dB, and each beam is centred on the relative beam axis.

**Table 1.** Main beam characteristics. $U_{peak}$ and $V_{peak}$ are computed with respect to the telescope line of sight. The Full Width (FW, minimum and maximum) at $-3$, $-10$, $-20$ dB are also reported since the main beams are strongly distorted by the optics, due to the off axis location of the feeds in the focal plane unit. The Full Width Half Maximum (FWHM) is the average value between the minimum and maximum of the beam width at $-3$ dB. The main beam directivity (Dir), the cross polar discrimination factor (XPD), and the main beam depolarisation ($d$) are reported, as computed in Appendix A.

| Feed Model | $U_{peak}$ | $V_{peak}$ | FW at $-3$dB (arcmin) | | FW at $-10$dB (arcmin) | | FW at $-20$dB (arcmin) | | FWHM (arcmin) | Dir (dBi) | XPD (dB) | $d$ (%) |
|---|---|---|---|---|---|---|---|---|---|---|---|---|
| | | | min | max | min | max | min | max | | | | |
| LFI4 A | 0.05142 | 0.03850 | 11.87 | 13.84 | 21.49 | 26.11 | 29.93 | 37.31 | 12.85 | 59.05 | 26.41 | 0.62 |
| LFI4 B | 0.05142 | 0.03851 | 11.63 | 13.01 | 21.76 | 25.01 | 29.75 | 36.54 | 12.32 | 59.14 | 24.75 | 0.95 |
| LFI4 C | 0.05143 | 0.03851 | 11.39 | 12.78 | 21.10 | 24.20 | 29.56 | 36.92 | 12.08 | 59.34 | 25.17 | 0.90 |
| LFI4 D | 0.05141 | 0.03852 | 11.63 | 13.21 | 21.10 | 25.13 | 29.27 | 36.24 | 12.42 | 59.28 | 24.91 | 0.93 |
| LFI9 A | -0.01767 | 0.05553 | 9.19 | 11.87 | 16.45 | 21.62 | 22.27 | 31.85 | 10.53 | 60.69 | 27.46 | 0.45 |
| LFI9 B | -0.01767 | 0.05552 | 8.88 | 11.14 | 15.56 | 19.86 | 20.96 | 33.14 | 10.01 | 61.13 | 27.23 | 0.56 |
| LFI9 C | -0.01767 | 0.05554 | 8.56 | 10.62 | 15.20 | 20.28 | 20.56 | 33.58 | 9.59 | 61.19 | 26.09 | 0.67 |

### 5.2. Intermediate beam

Intermediate beams have been computed in spherical polar cuts with $-5° < \theta_{bf} < 5°$ ($\Delta\theta = 0.1°$) and $0° < \phi_{bf} < 180°$ ($\Delta\phi = 0.5°$), using MrGTD up to the $2^{nd}$ order for all models of the LFI4 and for the 9A and 9C, and up to the $3^{rd}$ order for the 9B[4]. In Fig. 7 intermediate beams of the four models of LFI4 and the three models of LFI9 are shown, whereas, for clearly, in Fig. 8 the polar cuts at $\phi = 0°$ for the four models of the LFI4 and the three models of the LFI9 are reported. In the left panel, it would be noted the best performances of 4C respect to 4B and 4D, ET being equal. In the right panel, differences of about 6 dB between 9A and 9B, and 4 dB between 9B and 9C appear, reflecting the differences between horn edge tapers.

### 5.3. Far beam

Full pattern simulations have been carried out using MrGTD up to the $2^{nd}$ order for the 4B, 4C, and 4D, and up to the $3^{rd}$ order for the 4A, 9A, 9B, and 9C. Fig. 9 reports the antenna pattern response at large angles from the beam centre for the azimuthal cut corresponding to the main spillover, for all considered designs. As expected the radiation passing through the higher part of the primary reflector, the so-called main spillover, at about $\phi_{bf} = 50°$ and $\theta_{bf} = 90°$ for the LFI9 and at about $\phi_{bf} = 130°$ and $\theta_{bf} = 90°$ for the LFI4, raises increasing the illumination at the edge of the primary mirror ($-6$ dBi for 9A, $-2$ dBi for 9B, and about 1 dBi for 9C; $-14$ dBi for 4A and about $-2$ dBi for 4B, 4C, and 4D).

Resulting $4\pi$ maps of the best feed horn models analysed (4C and 9B), in terms of angular resolution and straylight rejection, are presented in Fig. 10 and 11, as function of the two polar coordinates $\theta_{bf} \in [-180°; +180°]$ and $\phi_{bf} \in [0°; +180°)$. The beam boresight direction is at the top of each map. Each polar cut with a constant $\phi_{bf}$ value is a meridian of the map. The $\theta_{bf}$ angle runs, on each meridian, from $\theta_{bf} = 0°$ (towards the main beam direction) to $\theta_{bf} = 180°$ ($-180°$) sweeping the map on its left (right) side. The $\phi_{bf}$ angle goes from $\phi_{bf} = 0°$ (centre of the map) to $\phi_{bf} = 180°$ (left side of the map) on the left side of the map, and from $\phi_{bf} = 0°$ (right side of the map) to $\theta_{bf} \to 180°$ (centre of the map) on the right side of the map. Therefore the centre of the map has $\theta_{bf} = 90°$ and $\phi_{bf} = 0°$. The $4\pi$ maps computed with the other feed models are reported in Appendix B.

---

[4] Not all beam patterns have been computed considering $3^{rd}$ order optical contributions since it takes about one month per beam. As reported in Sect. 6, we have found that neglecting $3^{rd}$ order optical contributions implies only few % errors on straylight evaluation.

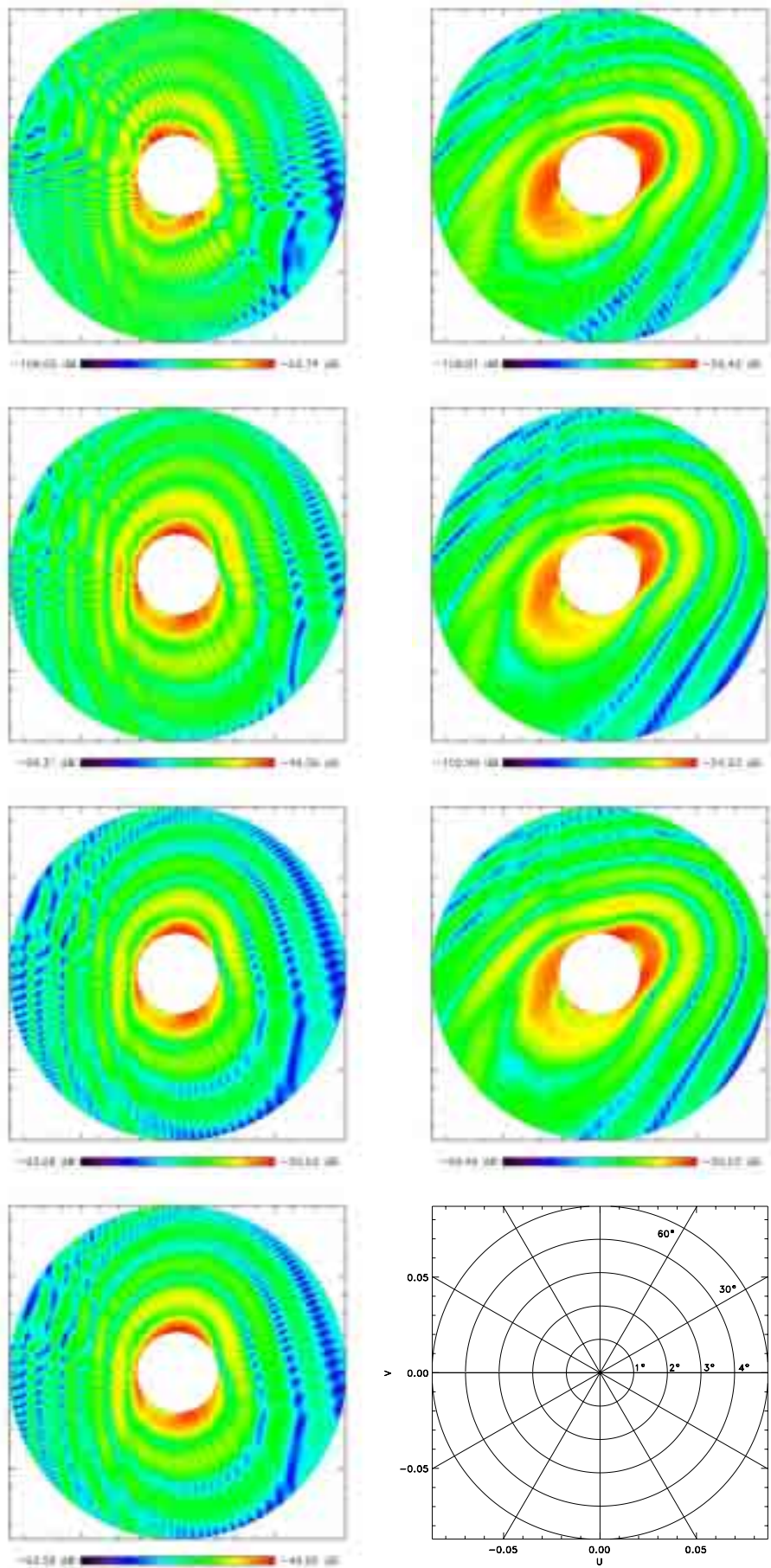

**Fig. 7.** Contour plots in the UV– plane (−0.087 < U,V < 0.087) of the intermediate beam computed for the four models of LFI4 (*left column*, from the top to the bottom: 4A, 4B, 4C, 4D) and for the three models of LFI9 (*right column*, from the top to the bottom: 9A, 9B, 9C).

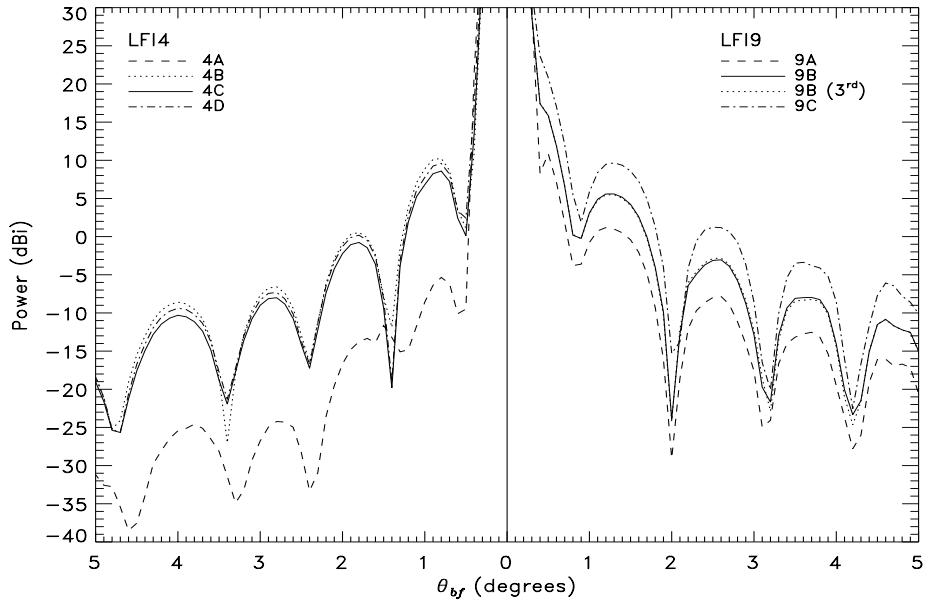

**Fig. 8.** Intermediate beams computed using the four feed models for the LFI4 and the three feed models for LFI9: 4A, 4B, 4C, 4D (*left panel*) and 9A, 9B, 9C (*right panel*), all computed using MrGTD up to the $2^{nd}$ order except for the 9B for which also the $3^{rd}$ order interaction has been simulated. Polar cuts at $\phi_{bf} = 0°$ are shown. In the left panel, it would be noted the best performances of 4C respect to 4B and 4D, ET being equal. In the right panel, differences of about 6 dB between 9A and 9B, and 4 dB between 9B and 9C appear, reflecting the differences between horn edge tapers.

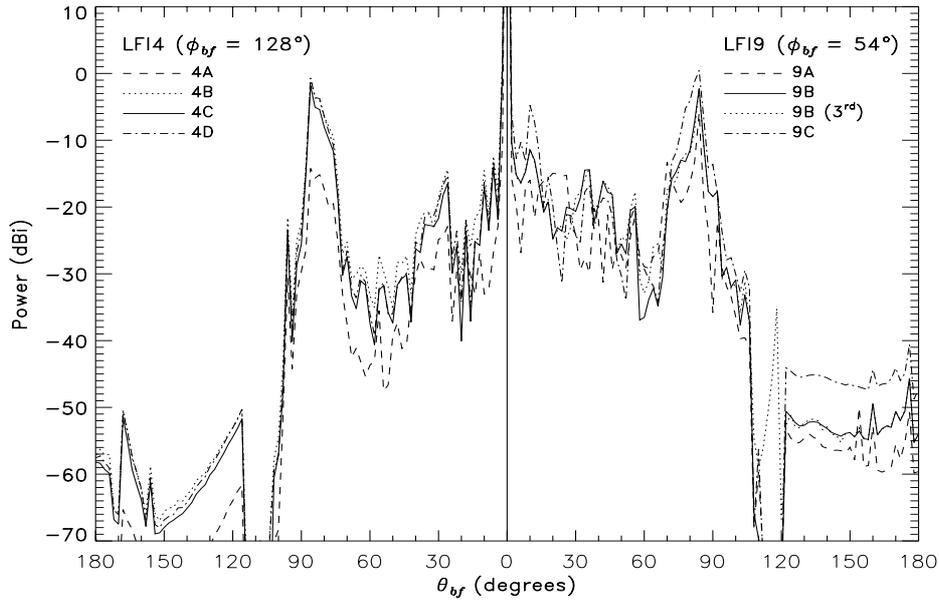

**Fig. 9.** Far-field antenna response for the azimuthal direction, reported in each panel, corresponding to the main spillover. Beams computed using the four feed models for the LFI4 and the three feed models for LFI9 are reported: 4A, 4B, 4C, 4D (*left panel*) and 9A, 9B, 9C (*right panel*), all computed using MrGTD up to the $2^{nd}$ order. In the case of LFI9 B we report the beam computed by including also the $3^{rd}$ order optical interaction, for comparison.

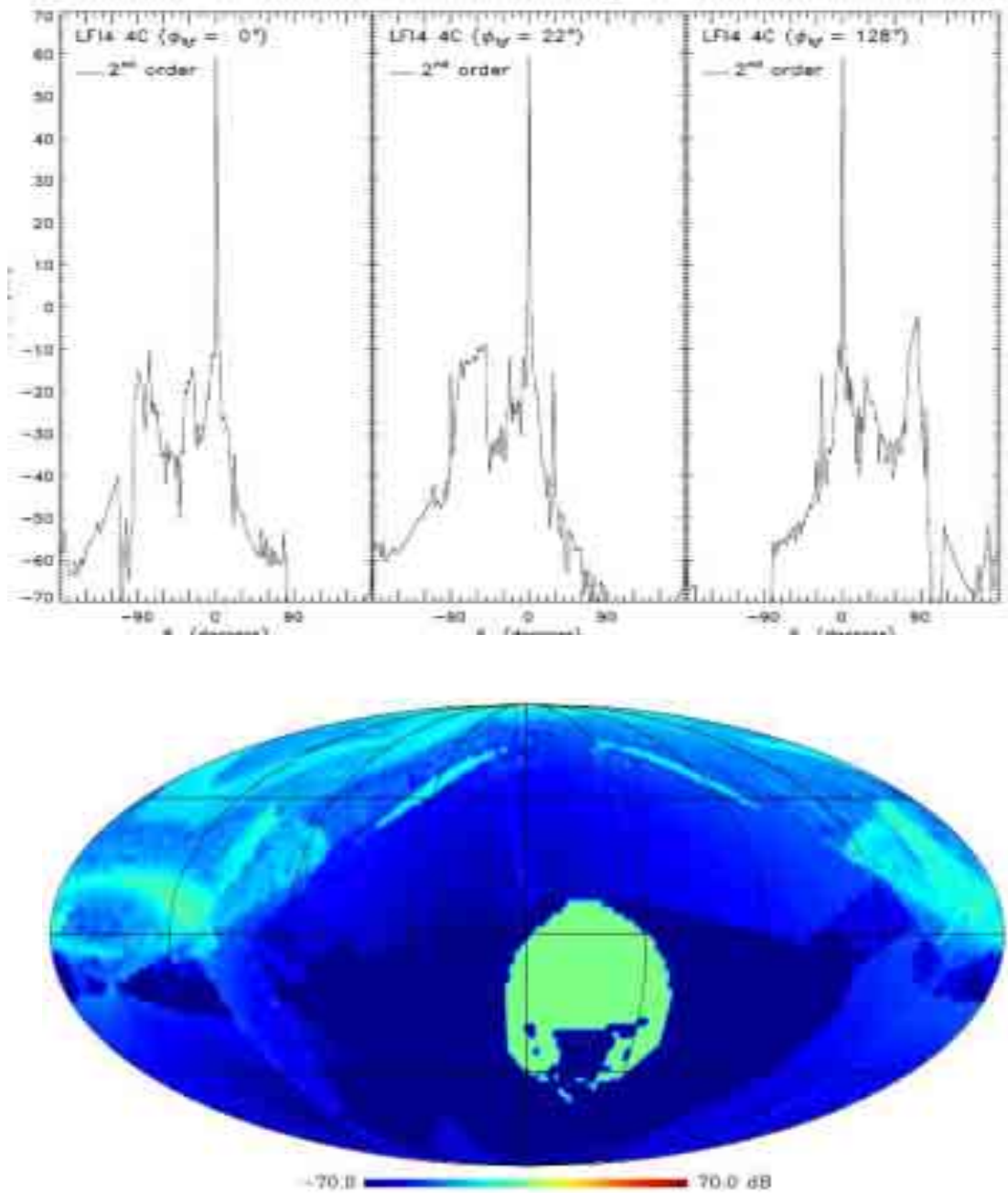

**Fig. 10.** Full pattern of the LFI4 C at 100 GHz, computed with the MrGTD up to the $2^{nd}$ order, except in the main beam region that has been computed using the PO/PTD analysis in order to avoid caustic artefacts originated by the GTD approach. In the upper side of the figure, three polar cuts are reported: $\phi_{bf}$ equal to 0° (*left panel*), $\phi_{bf}$ equal to 22° (*centre panel* – cut passing through one of the two wings originated by the rays reflected on the inner part of the baffle), and $\phi_{bf}$ equal to 128° (*right panel* – cut passing through the main spillover). At the bottom, the $4\pi$ map is shown. In the uniform region around $\theta_{bf} \simeq -100°$ and $\phi_{bf} \simeq 157.5°$ (artfully settled to zero dB level) no ray coming from the source has been found and thus the total field is null. The mollweide projection of the beam has been obtained with the HEALPix package (Górski et al. 1998, see also HEALPix home page at http://www.eso.org/science/healpix/).

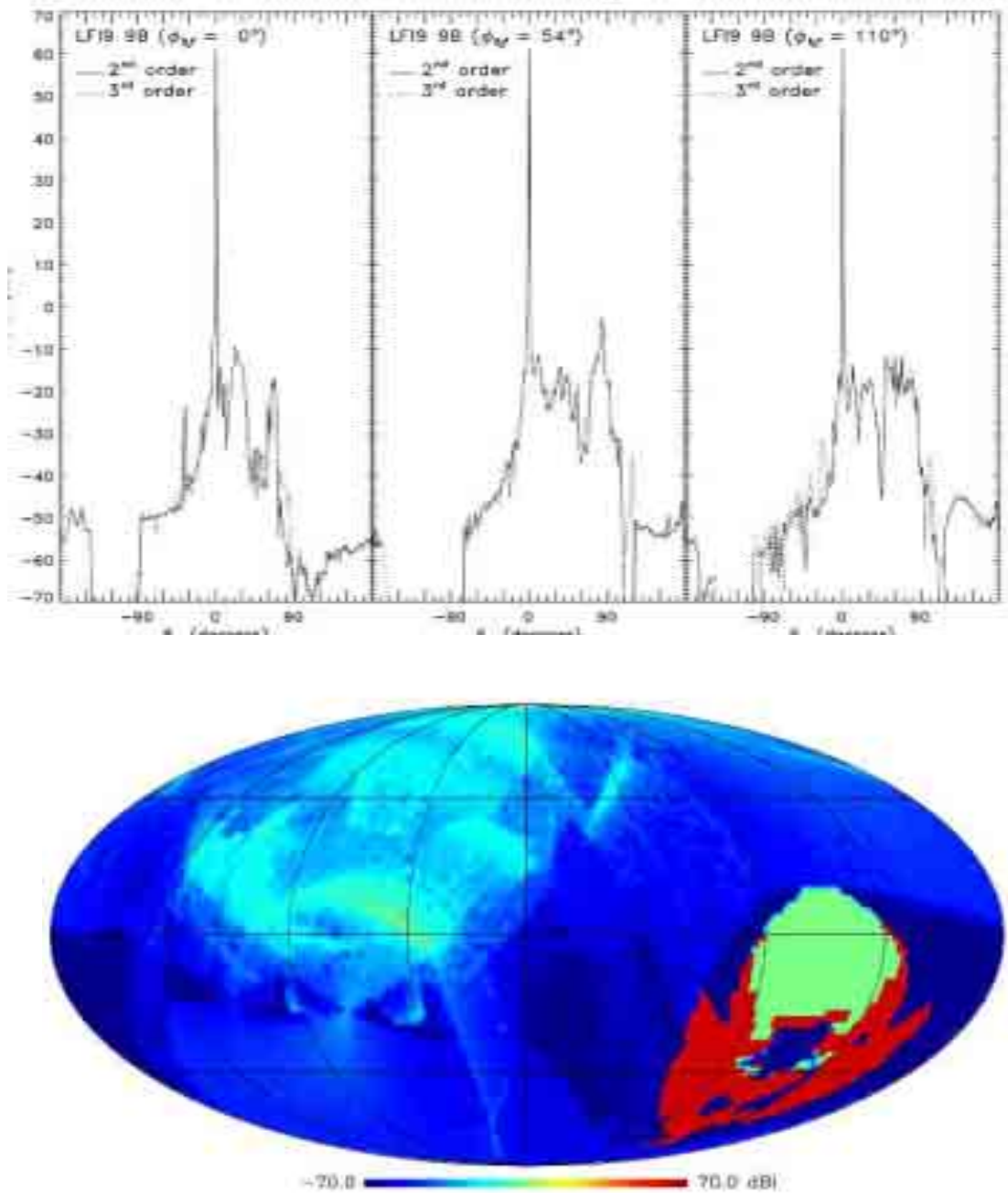

**Fig. 11.** Full pattern of the LFI9 B at 100 GHz, computed with the MrGTD up to the $2^{nd}$ order, except in the main beam region that has been computed using the PO/PTD analysis in order to avoid caustic artefacts originated by the GTD approach. In the upper side of the figure, three polar cuts are reported: $\phi_{bf}$ equal to 0° (*left panel*), $\phi_{bf}$ equal to 54° (*centre panel* – cut passing through the main spillover), and $\phi_{bf}$ equal to 110° (*right panel* – cut passing through one of the two wings originated by the rays reflected on the inner part of the baffle). At the bottom, the $4\pi$ map is shown. In the uniform region around $\theta_{bf} \simeq -100°$ and $\phi_{bf} \simeq 67.5°$ (artfully settled to zero dB level) no ray coming from the source has been found and thus the total field is null. With respect to the other maps of full pattern, we have pointed out in this case the region immediately below (artfully settled to 70 dB level), where only the third order optical interaction gives no null contributions. See also Fig. 12.

In Fig. 12 is reported the map of the difference (in dB) between the full antenna pattern response computed by including $1^{st}$, $2^{nd}$ and $3^{rd}$ order optical interactions and that obtained neglecting the $3^{rd}$ order optical interaction, in the beam reference frame and for the model B of the LFI9. The contribution from the $3^{rd}$ order optical interactions would seems not negligible in the far sidelobes.

Since the computation of high order optical interactions is very time consuming it is useful for optical optimisation studies to quantify the loss of accuracy in the straylight contamination evaluation introduced by neglecting the $3^{rd}$ order optical interactions, and this has been done in Sect. 6.

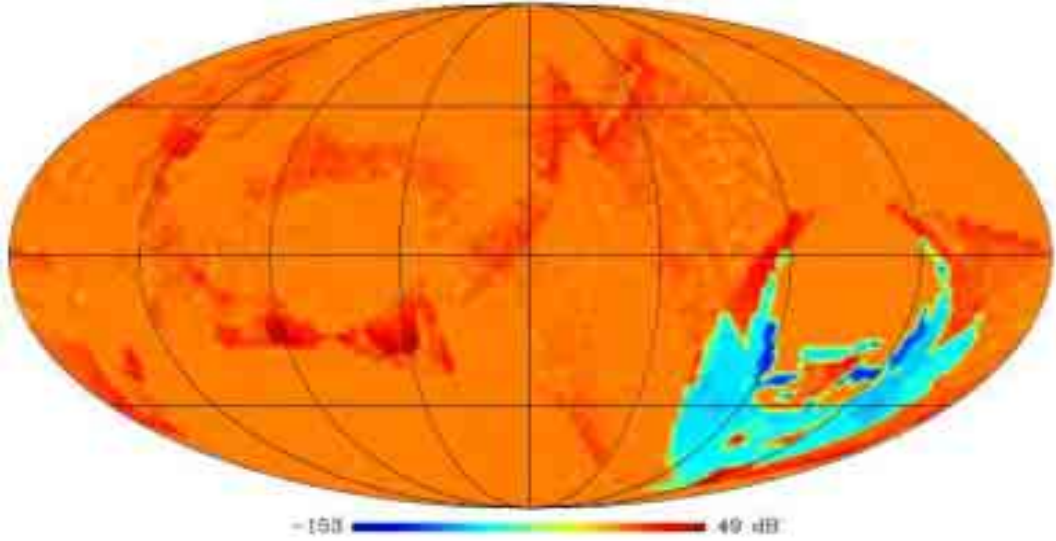

**Fig. 12.** Map of the difference (in dB) between the full antenna pattern response computed by including $1^{st}$, $2^{nd}$ and $3^{rd}$ order optical interactions and that obtained neglecting the $3^{rd}$ order optical interaction, in the beam reference frame and for the model B of the LFI9. The uniform region at about $\theta_{bf} \sim -100°$, $\phi_{bf} \sim 67.5°$ is where the contributions up to the $3^{rd}$ order optical interaction is null, while the region immediately below is where only the $3^{rd}$ order optical interaction gives no null contributions. Therefore, the reported values in the latter region are equal to the full antenna pattern response (in dBi) due to the $3^{rd}$ optical interaction alone (see also bottom panel of Fig. 11). It is also interesting to note the *spot* at $\theta_{bf} \sim 115°$, $\phi_{bf} \sim 54°$ corresponding to the significant contribution from the $3^{rd}$ order optical interaction clearly visible in the right panel of Fig. 9 (dotted line).

## 6. Discussion

In Tab. 2 are summarised the fractional contributions to the integrated antenna pattern, $f_\%$, in each considered pattern region, $\Omega$, defined as $100 \int_\Omega J d\Omega / \int_{4\pi} d\Omega$, for each feed horn model analysed. It is evident the good level of the pattern convergence from $2^{nd}$ to $3^{rd}$ order contributions, since only differences at 0.1% appear in $f_\%$. The Galactic straylight contamination (in $\mu$K antenna temperature, RMS and peak-to-peak), as computed in Paper II considering contributions from dust, diffuse free-free emission, diffuse synchrotron emission, and HII regions, are also reported for each pattern region. Differences in $f_\%$, derived from different feed horn designs and hence different main reflector illuminations, involve differences of about 0.3% in the Galactic straylight contamination. It is noteworthy that neglecting $3^{rd}$ order optical contributions implies only few % errors on the straylight evaluation. This is very important in an optimisation activity since tell us that the computation accuracy of an optical simulation carried out using the MrGTD up to the $2^{nd}$ order is sufficient to provide robust straylight results, saving about 75% of computational time (about one week for a full pattern simulation with MrGTD up to the $2^{nd}$ order against about one month for a full pattern simulation with MrGTD up to the $3^{td}$ order, using a 550 MHz Pentium II machine).

In Fig. 13 $f_\%$ versus the LET is shown, as computed for the four models of the LFI4 and the three models of the LFI9 at 100 GHz. Dashed, dotted, and solid lines are the fitted curves (forcing the obvious zero crossing) of these points for the intermediate beam computed up to the $2^{nd}$ order, the full pattern computed up to the $2^{nd}$ order, and, only for the LFI9, the full pattern computed up to the $3^{rd}$ order, respectively.

**Table 2.** Intermediate and far beams integrated power, normalised to the total power ($f\%$), for each simulated feed horn. Galactic straylight contamination (GSC, in $\mu$K RMS and peak-to-peak antenna temperature) computed in Paper II considering contributions from dust, diffuse free-free emission, diffuse synchrotron emission, and HII regions, are also reported for each pattern region. For the far beam region, when the $3^{rd}$ order MrGTD contribution has been computed, the corresponding GSC has been reported (n.a. means *not available*).

| Feed Model | Edge Taper dB @ 24° | Intermediate Beam | | | Far Beam ($3^{rd}$ order) | | | Far beam ($2^{nd}$ order) | | |
|---|---|---|---|---|---|---|---|---|---|---|
| | | f% | $GSC_{RMS}$ | $GSC_{p2p}$ | f% | $GSC_{RMS}$ | $GSC_{p2p}$ | f% | $GSC_{RMS}$ | $GSC_{p2p}$ |
| LFI9 A | 25.5 | 0.010 | $8.38\times10^{-3}$ | $2.94\times10^{-1}$ | 0.185 | $7.56\times10^{-2}$ | $6.38\times10^{-1}$ | 0.175 | $7.33\times10^{-2}$ | $6.16\times10^{-1}$ |
| LFI9 B | 19.0 | 0.034 | $2.78\times10^{-2}$ | $8.84\times10^{-1}$ | 0.378 | $1.99\times10^{-1}$ | $1.92\times10^{+0}$ | 0.351 | $1.90\times10^{-1}$ | $1.87\times10^{+0}$ |
| LFI9 C | 15.0 | 0.099 | $8.08\times10^{-2}$ | $2.26\times10^{+0}$ | 1.186 | $7.05\times10^{-1}$ | $6.35\times10^{+0}$ | 1.086 | $6.72\times10^{-1}$ | $6.18\times10^{+0}$ |
| LFI4 A | 28.3 | 0.002 | $1.86\times10^{-3}$ | $6.51\times10^{-2}$ | 0.041 | $1.48\times10^{-2}$ | $1.10\times10^{-1}$ | 0.036 | $1.36\times10^{-2}$ | $1.11\times10^{-1}$ |
| LFI4 B | 19.0 | 0.057 | $4.88\times10^{-2}$ | $1.65\times10^{+0}$ | n.a. | n.a. | n.a. | 0.459 | $2.09\times10^{-1}$ | $1.63\times10^{+0}$ |
| LFI4 C | 19.0 | 0.042 | $3.67\times10^{-2}$ | $1.27\times10^{+0}$ | n.a. | n.a. | n.a. | 0.292 | $1.40\times10^{-1}$ | $1.12\times10^{+0}$ |
| LFI4 D | 19.0 | 0.053 | $4.57\times10^{-2}$ | $1.58\times10^{+0}$ | n.a. | n.a. | n.a. | 0.398 | $1.92\times10^{-1}$ | $1.56\times10^{+0}$ |

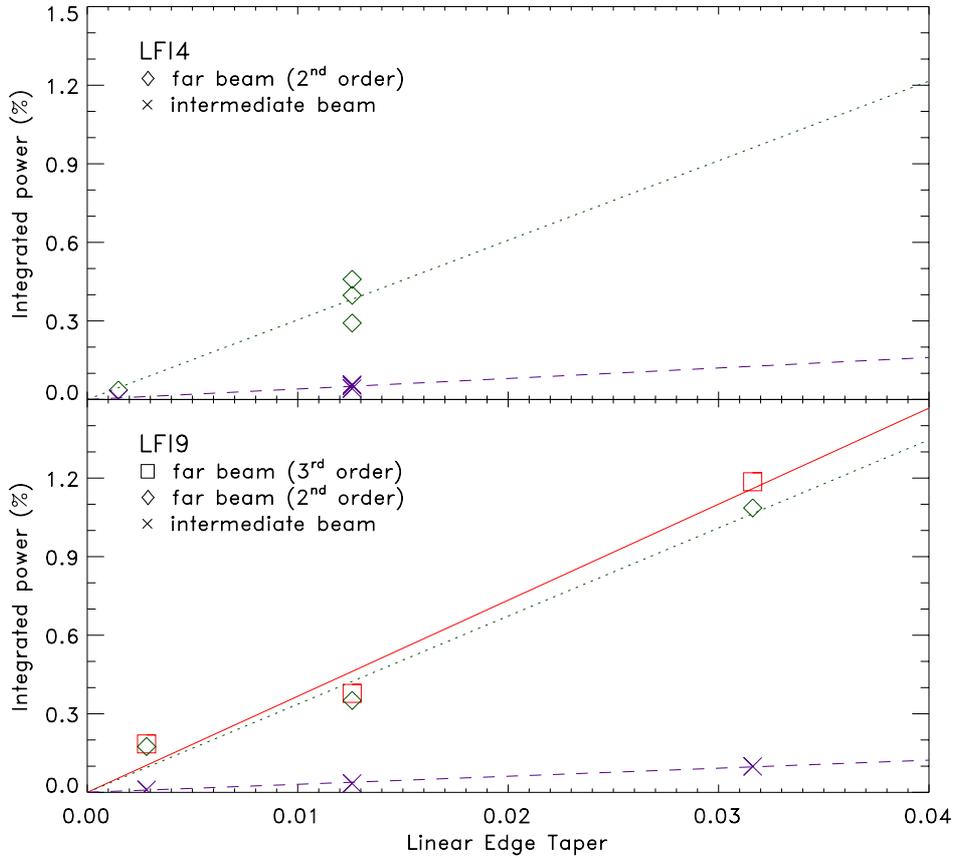

**Fig. 13.** Fractional contribution (in per cent) to the integrated antenna pattern as computed for the four models of LFI4 and the three models of LFI9 at 100 GHz. Dashed, dotted, and solid lines are the fitted curves (forcing the obvious zero crossing) of these points for the intermediate beam computed up to the $2^{nd}$ order, the full pattern computed up to the $2^{nd}$ order, and, only for the feed horn #9, the full pattern computed up to the $3^{td}$ order, respectively. For the LFI4 the fitted angular coefficients are 30.40 and 4.01 in the far and intermediate beam, respectively; whereas for the LFI9 they are 36.68 for the far beam computed up to $3^{rd}$ order, 33.65 for the computation up to the $2^{nd}$ order, and 3.08 for the intermediate beam.

Finally, we have found linear approximations describing the dependence of the fractional contribution to the integrated antenna pattern, $f_\%$, by the LET, from the considered pattern regions:

$$f_\% \simeq 3.08 \times \text{LET} \tag{1}$$

for the intermediate beam region, and:

$$f_\% \simeq 36.68 \times \text{LET} \qquad \text{considering } 3^{rd} \text{ order optical contributions} \qquad (2)$$

$$f_\% \simeq 33.65 \times \text{LET} \qquad \text{considering } 2^{nd} \text{ order optical contributions} \qquad (3)$$

for the far beam region.

These equations concern the LFI9 at 100 GHz, for which the simulated feed models had three different edge taper. The same has been done for the LFI4 and the resulting relations are

$$f_\% \simeq 4.01 \times \text{LET} \qquad (4)$$

for the intermediate beam region, and:

$$f_\% \simeq 30.40 \times \text{LET} \qquad \text{considering } 2^{nd} \text{ order optical contributions} \qquad (5)$$

for the far sidelobes.

The fractional contribution to the integrated antenna pattern has been related to the Galactic straylight contamination in Paper II (see Equations 3–6). Fig. 14 shows the straylight contamination (peak-to-peak and RMS values of the antenna temperature in $\mu$K) versus the LET, as computed for the four models of the LFI4 and the three models of the LFI9 at 100 GHz. Contamination due to the intermediate beam, far sidelobes and the total value are reported separately.

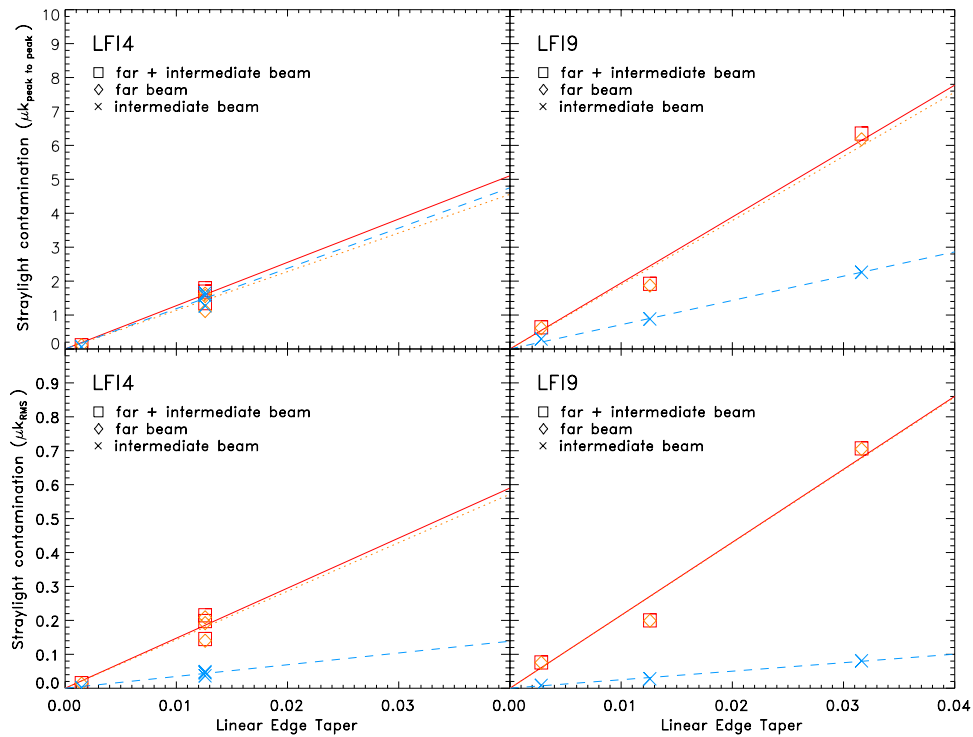

**Fig. 14.** Correlation between Galactic straylight contamination and linear edge taper for the 100 GHz feed horn #4 and #9. Solid lines are the fitted curve (forcing the obvious zero crossing) of the four points for LFI4 and of the three points for LFI9, whose angular coefficients represent the straylight degradation factor (127.64 $\mu$K peak-to-peak/LET and 14.75 $\mu$K RMS /LET for the LFI4; 194.42 $\mu$K peak-to-peak /LET and 21.51 $\mu$K RMS /LET for the LFI9) due to far side lobes and intermediate beam. Dashed lines represent the fitted curve of the four points for LFI4 and of the three points for LFI9 due to intermediate beams (118.81 $\mu$K peak-to-peak /LET and 3.46 $\mu$K RMS /LET for the LFI4; 71.52 $\mu$K peak-to-peak /LET and 2.51 $\mu$K RMS /LET for the LFI9).

## 7. Conclusions

One of the major sources of systematic errors in CMB experiments is the non ideal response of the optics. Main beam aberrations and sidelobes may degrade the reconstruction of the power spectrum of the CMB anisotropies at high and low multipoles, respectively. While for the main beam region the Physical Optics gives accurate and well established results taking into account only the feed horn and the telescope alone, for the intermediate and sidelobe regions, the optics shall include also the structures surrounding the telescope. This complicates the electromagnetic model increasing the computational time and may preclude the ability to perform full pattern calculations. On the other hand for a satellite mission like PLANCK the knowledge of the full pattern response is a decisive factor for guarantee the high performances of the instruments onboard (LFI and HFI). For this reason we studied in detail the way for predicting the radiation pattern at LFI frequencies using GRASP8.

The electromagnetic model of PLANCK LFI has been setup. The model includes the feed pattern, the telescope optics, the first V– groove and the shield surrounding the telescope. Moreover, for this geometry, the sequences of reflections/diffractions which give a significant contribution (> 100 dB below the beam peak) has been identified. This assures a reasonable full pattern extimation for robust straylight evaluation. It has been found that the contribution of more than 2 optical interactions is not required in the optimization activity, since the neglecting $3^{rd}$ order optical contributions involve differences less than 3% on the straylight evaluation, saving about 75% of the computational time.

The inadequacy of a pure Gaussian feed model in realistic far beam predictions is demonstrated: relevant features in the full pattern are also due to the sidelobes in the feed horn pattern. Not only the realistic pattern needs to be considered but also the detail of the corrugation design could impacts the beam characteristics. Different corrugation profiles, edge taper being equal, involves differences of about 3% in the main beam angular resolution and about 40% on the straylight signal, although remains below 2 $\mu$K in all cases of the LFI4. The 4C model (with an angular resolution of $10'$ and a straylight contamination of 1.12 $\mu$K peak-to-peak) and the 9B model (with an angular resolution of $12'$ and a straylight contamination of 1.87 $\mu$K peak-to-peak) models represent the best choice for the LFI feed horn #4 and #9, reaching the goal and the requirement, respectively.

Linear fit of straylight contamination (peak-to-peak and rms values) and Linear Edge Taper has been obtained. Although these fits have been specifically derived from the channel studied here, could be used to set at the first order the ET of the other LFI channels with feed horns located near those studied, without any pattern simulations. However, sofisticated pattern simulations are mandatory to accurately quantify the beam aberrations and the straylight contamination.

*Acknowledgements.* We wish to thank people of the Herschel/PLANCK Project, ALCATEL Space Industries, and the LFI Consortium that are involved in activities related to optical simulations. Some of the results in this paper have been derived using the HEALPix (Górski, Hivon, and Wandelt 1999). Thanks to Per Nielsen (TICRA Engineering Consultants, Copenhagen), for his clarifying correspondence about MrGTD and support with GRASP8.

## Appendix A: Main beam characteristics

The directivity of the co– polar ($D_{cp}$) and x– polar ($D_{xp}$) components have been computed using GRASP8. Then the cross polar discrimination factor (XPD) has been defined as:

$$XPD = \frac{D_{cp}}{D_{xp}} \tag{A.1}$$

The depolarization parameter has been obtained computing the Stokes parameters in each point of the regular UV– grid, for each beam.

$$S_I(u,v) = E_{cp}(u,v)^2 + E_{xp}(u,v)^2 \tag{A.2}$$
$$S_Q(u,v) = E_{cp}(u,v)^2 - E_{xp}(u,v)^2 \tag{A.3}$$
$$S_U(u,v) = 2 \cdot E_{cp}(u,v) \cdot E_{xp}(u,v) \cdot \cos[\delta\varphi(u,v)] \tag{A.4}$$
$$S_V(u,v) = 2 \cdot E_{cp}(u,v) \cdot E_{xp}(u,v) \cdot \sin[\delta\varphi(u,v)] \tag{A.5}$$

in which $E_{cp}$ and $E_{xp}$ are the amplitude field of the co– polar and x– polar components, respectively, and $\delta\varphi$ is the phase difference between the co– polar and x– polar fields.

Then, over the whole UV– plane computed, each parameter has been summed.

$$S_N = \sum_{(u,v)} S_N(u,v) \cdot \Delta u \Delta v \qquad \text{where N=I,Q,U,V} \tag{A.6}$$

and, finally,

$$d(\&) = \left(1 - \frac{\sqrt{S_Q^2 + S_U^2 + S_V^2}}{S_I}\right) \cdot 100. \tag{A.7}$$

## Appendix B: Far beam patterns

In the follows we report the far beam computed using the MrGTD up to the $2^{nd}$ order for the 4A, 4B, and 4D, and up to the $3^{rd}$ order for the 9A and 9C. In the upper side of each figure, three polar cuts are reported (the cut passing through the main spillover is included). At the bottom of each figure, the $4\pi$ map is shown and the previous cuts may be easily identify. As for the $4\pi$ maps shown in Sect. 5.3 the beam boresight direction is at the top of each map. Each polar cut with a constant $\phi_{bf}$ value is a meridian of the map. The $\theta_{bf}$ angle runs, on each meridian, from $\theta_{bf} = 0°$ (towards the main beam direction) to $\theta_{bf} = 180°$ ($-180°$) sweeping the map on its left (right) side. The $\phi_{bf}$ angle goes from $\phi_{bf} = 0°$ (centre of the map) to $\phi_{bf} = 180°$ (left side of the map) on the left side of the map, and from $\phi_{bf} = 0°$ (right side of the map) to $\theta_{bf} \rightarrow 180°$ (centre of the map) on the right side of the map. Therefore the centre of the map has $\theta_{bf} = 90°$ and $\phi_{bf} = 0°$.

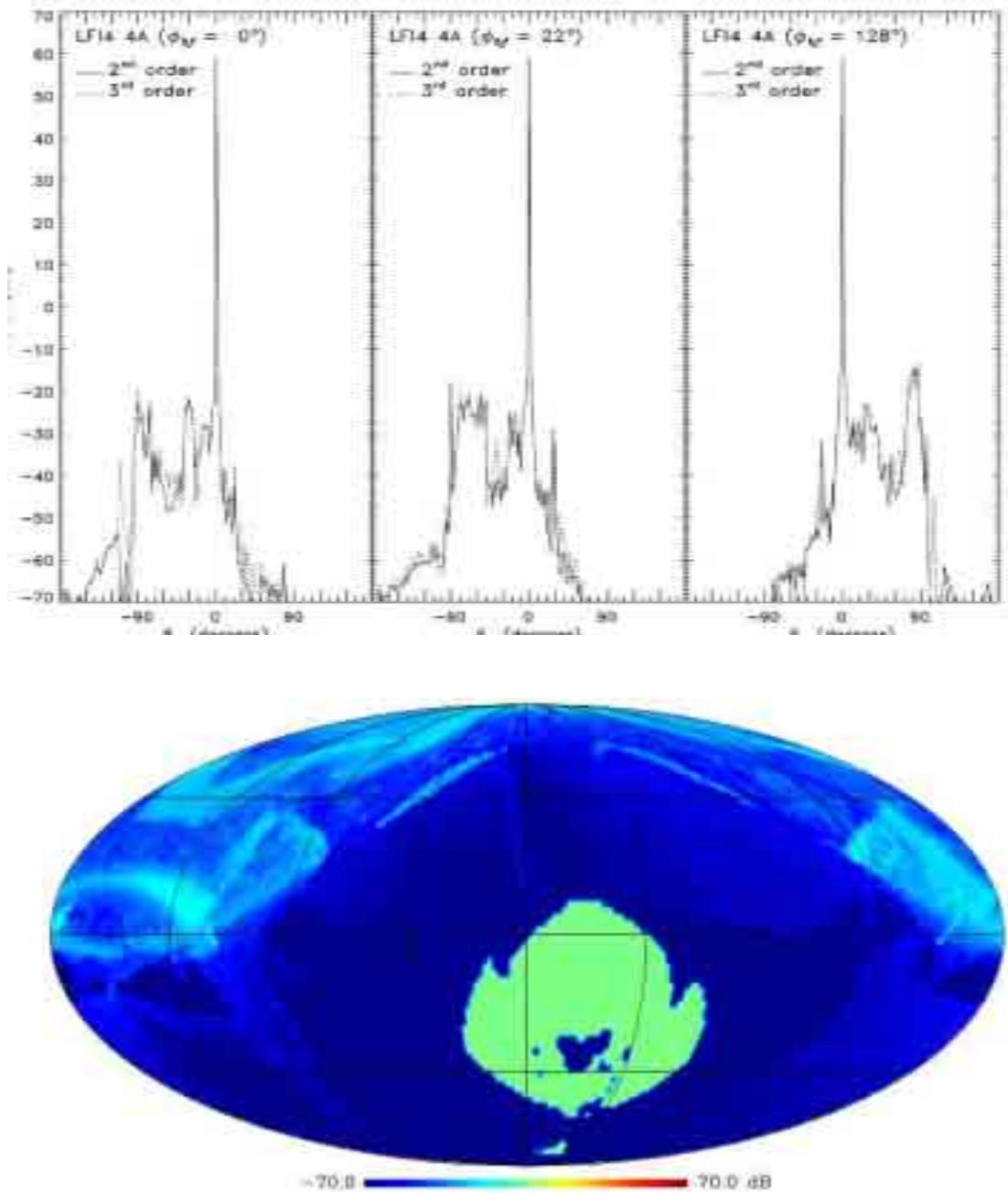

**Fig. B.1.** Full pattern of the LFI4 A at 100 GHz, computed with the MrGTD up to the $2^{nd}$ order, except in the main beam region that has been computed using the PO/PTD analysis in order to avoid caustic artefacts originated by the GTD approach. In the upper side of the figure, three polar cuts are reported: $\phi_{bf}$ equal to 0° (*left panel*), $\phi_{bf}$ equal to 22° (*centre panel* – cut passing through one of the two wings originated by the rays reflected on the inner part of the baffle), and $\phi_{bf}$ equal to 128° (*right panel* – cut passing through the main spillover). At the bottom, the $4\pi$ map is shown. In the uniform region around $\theta_{bf} \simeq -100°$ and $\phi_{bf} \simeq 157.5°$ (artfully settled to zero dB level) no ray coming from the source has been found and thus the total field is null.

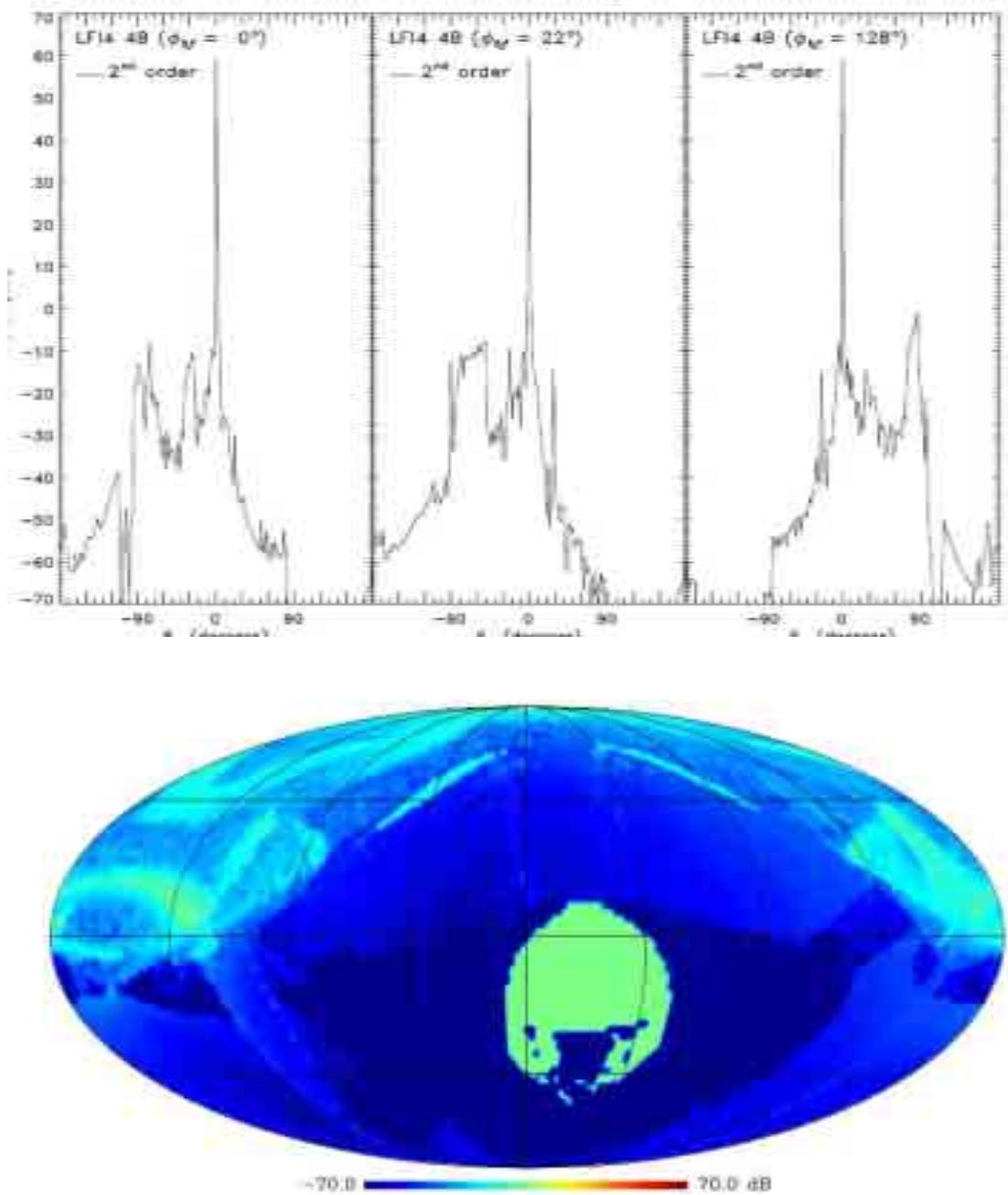

**Fig. B.2.** Full pattern of the LFI4 B at 100 GHz, computed with the MrGTD up to the $2^{nd}$ order, except in the main beam region that has been computed using the PO/PTD analysis in order to avoid caustic artefacts originated by the GTD approach. In the upper side of the figure, three polar cuts are reported: $\phi_{bf}$ equal to 0° (*left panel*), $\phi_{bf}$ equal to 22° (*centre panel* – cut passing through one of the two wings originated by the rays reflected on the inner part of the baffle), and $\phi_{bf}$ equal to 128° (*right panel* – cut passing through the main spillover). At the bottom, the $4\pi$ map is shown. In the uniform region around $\theta_{bf} \simeq -100°$ and $\phi_{bf} \simeq 157.5°$ (artfully settled to zero dB level) no ray coming from the source has been found and thus the total field is null.

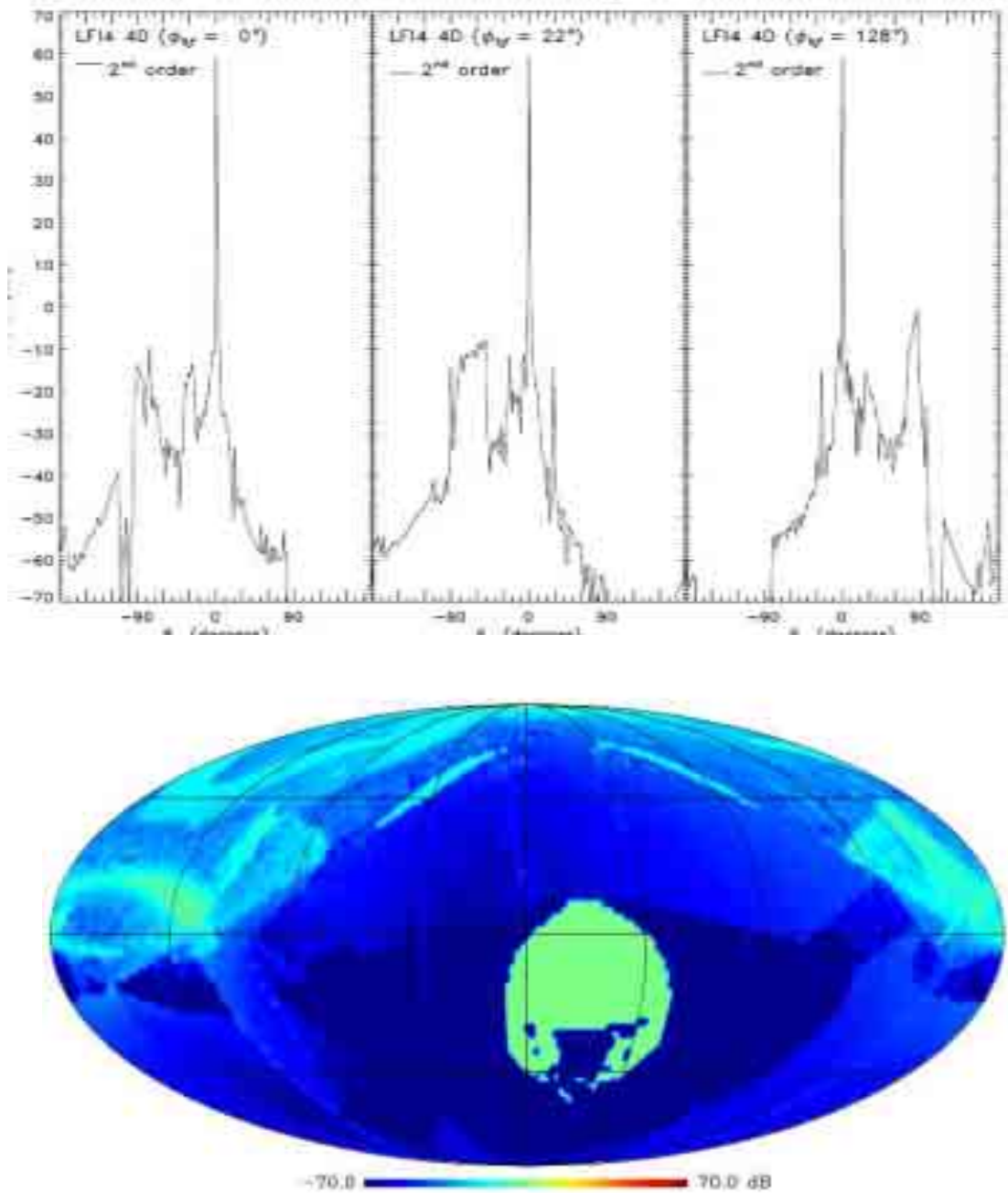

**Fig. B.3.** Full pattern of the LFI4 D at 100 GHz, computed with the MrGTD up to the $2^{nd}$ order, except in the main beam region that has been computed using the PO/PTD analysis in order to avoid caustic artefacts originated by the GTD approach. In the upper side of the figure, three polar cuts are reported: $\phi_{bf}$ equal to 0° (*left panel*), $\phi_{bf}$ equal to 22° (*centre panel* – cut passing through one of the two wings originated by the rays reflected on the inner part of the baffle), and $\phi_{bf}$ equal to 128° (*right panel* – cut passing through the main spillover). At the bottom, the $4\pi$ map is shown. In the uniform region around $\theta_{bf} \simeq -100°$ and $\phi_{bf} \simeq 157.5°$ (artfully settled to zero dB level) no ray coming from the source has been found and thus the total field is null.

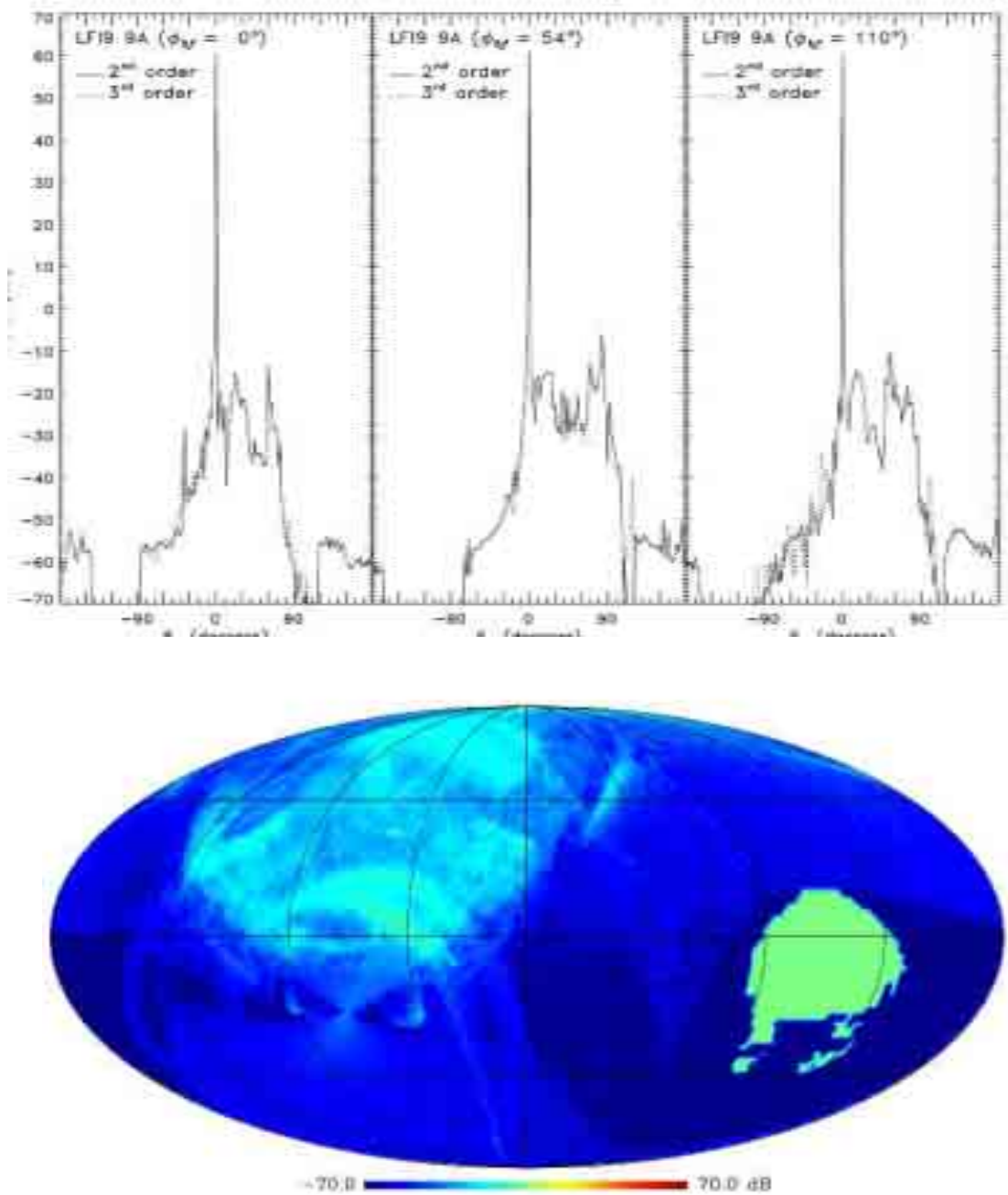

**Fig. B.4.** Full pattern of the LFI9 A at 100 GHz, computed with the MrGTD up to the $2^{nd}$ order, except in the main beam region that has been computed using the PO/PTD analysis in order to avoid caustic artefacts originated by the GTD approach. In the upper side of the figure, three polar cuts are reported: $\phi_{bf}$ equal to 0° (*left panel*), $\phi_{bf}$ equal to 54° (*centre panel* – cut passing through the main spillover), and $\phi_{bf}$ equal to 110° (*right panel* – cut passing through one of the two wings originated by the rays reflected on the inner part of the baffle). At the bottom, the $4\pi$ map is shown. In the uniform region around $\theta_{bf} \simeq -100°$ and $\phi_{bf} \simeq 67.5°$ (artfully settled to zero dB level) no ray coming from the source has been found and thus the total field is null.

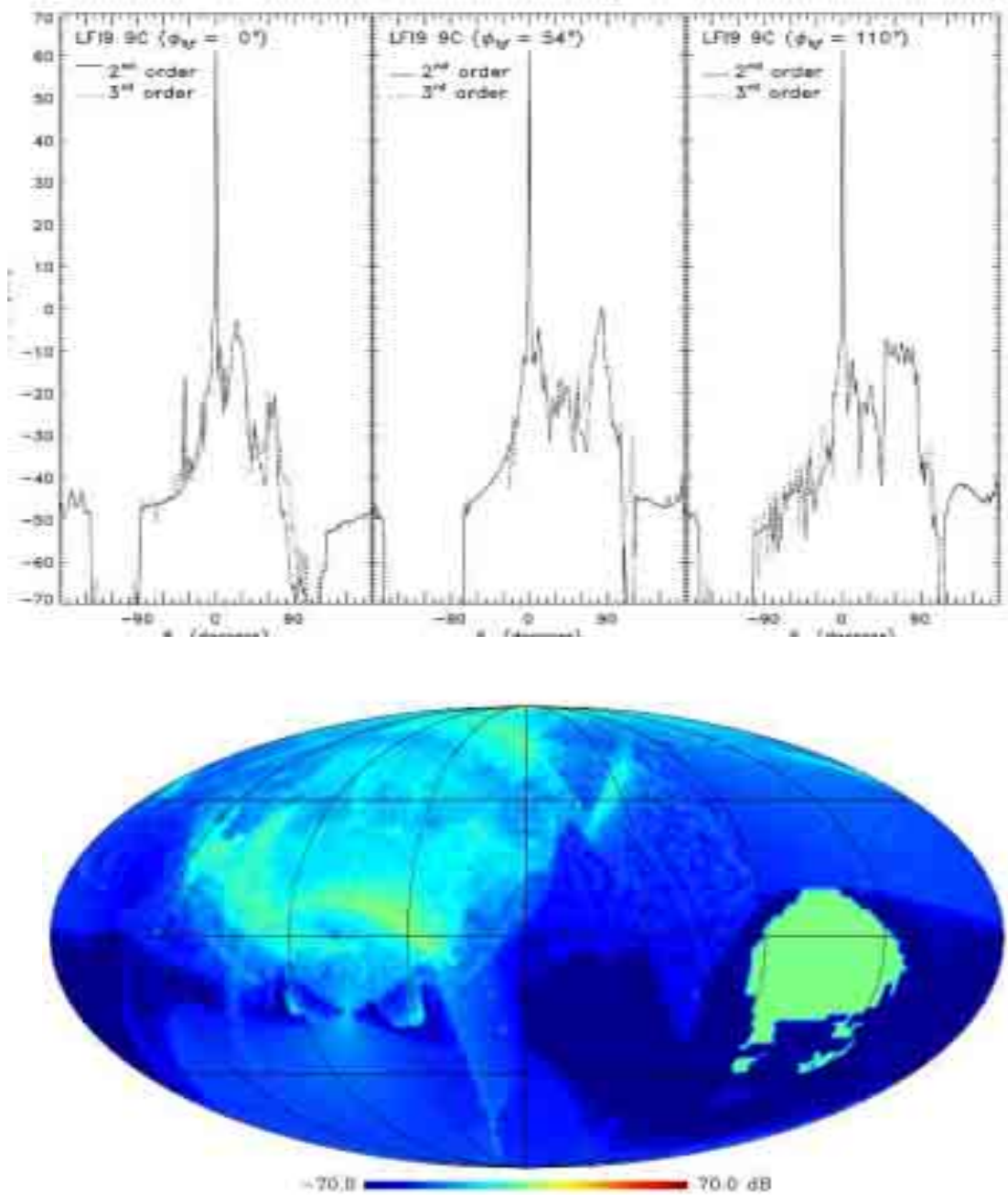

**Fig. B.5.** Full pattern of the LFI9 C at 100 GHz, computed with the MrGTD up to the $2^{nd}$ order, except in the main beam region that has been computed using the PO/PTD analysis in order to avoid caustic artefacts originated by the GTD approach. In the upper side of the figure, three polar cuts are reported: $\phi_{bf}$ equal to 0° (*left panel*), $\phi_{bf}$ equal to 54° (*centre panel* – cut passing through the main spillover), and $\phi_{bf}$ equal to 110° (*right panel* – cut passing through one of the two wings originated by the rays reflected on the inner part of the baffle). At the bottom, the $4\pi$ map is shown. In the uniform region around $\theta_{bf} \simeq -100°$ and $\phi_{bf} \simeq 67.5°$ (artfully settled to zero dB level) no ray coming from the source has been found and thus the total field is null.